# Free-sustaining Three-dimensional S235 Steel-based Porous Electrocatalyst for Highly Efficient and Durable Oxygen Evolution

Weijia Han[1], Karsten Kuepper[2], Peilong Hou[1], Wajiha Akram[1], Henning Eickmeier[1], Jörg Hardege[3], Martin Steinhart[1], and Helmut Schäfer[*1]

*1. Institute of Chemistry of New Materials and Center of Physics and Chemistry of New Materials, Universität Osnabrück, Barbarastrasse 7, 49076 Osnabrück, Germany*

*2. Department of Physics, Universität Osnabrück, Barbarastrasse 7, 49076 Osnabrück, Germany*

*3. School of Environmental Sciences, Hull University, HU67RX, Hull, U.K.*

*\* corresponding author*

# ABSTRACT

A novel oxygen evolution reaction (OER) catalyst (3D S235-P steel) based on steel S235 substrate has been successfully prepared via a facile one-step surface modification. The standard Carbon Manganese steel was phosphorized superficially leading to the formation of a unique 3D interconnected nanoporous surface with high specific area which facilitates the electrocatalytically initiated oxygen evolution reaction. The prepared 3D S235-P steel exhibits enhanced electrocatalytic OER activities in alkaline regime confirmed by a low overpotential ($\eta$=326 mV at j=10 mA cm$^{-2}$) and a small Tafel slope of 68.7 mV dec$^{-1}$. Moreover, the catalyst was found to be stable under long-term usage conditions functioning as oxygen evolving electrode at pH 13 as evidenced by the sufficient charge to oxygen conversion rate (Faradaic efficiency: 82.11% and 88.34% at 10 mA cm$^{-2}$ and 5 mA cm$^{-2}$, respectively). In addition, it turned out that the chosen surface modification renders steel S235 into an OER electrocatalyst sufficiently and stable to work in neutral pH condition. Our investigation revealed that the high catalytic activities are likely to stem from the generated Fe/(Mn) hydroxide/oxo-hydroxides generated during the OER process. The phosphorization treatment is therefore not only an efficient way to optimize the electrocatalytic performance of standard Carbon-Manganese steel, but also enables for the development of low cost and abundant steels in the field of energy conversion.

Because of the overexploitation of non-renewable fossil energy sources and the related environmental pollution, energy shortage issues have become an urgent and serious phenomenon, stimulating a sharp increase in the search for alternative, green energies utilizing a large variety of potentially possible methods [1, 2, 3, 4, 5]. Electrochemical splitting of water, as widely accepted, is one of the most expedient and environmentally friendly techniques for a possible clean hydrogen energy production [6, 7, 8, 9, 10, 11, 12, 13, 14, 15, 16, 17, 18, 19, 20, 21, 22, 23, 24, 25]. However, the efficiency of this method is mainly restricted by the sluggish anodic oxygen evolution reaction (OER), the rate-limiting step for overall water splitting, which involves four sequential proton-coupled electron transfer steps and the formation of an oxygen–oxygen bonding [26, 27, 28, 29]. Generally, precious metal oxides like $IrO_2$ and $RuO_2$ are seen as benchmark OER catalysts due to low overpotentials especially caused on the anode side [30, 31]. However, scarcity, high acquisition costs and poor performance in long-term OER measurements in alkaline solution hamper the suitability for widespread practical applications [32,33]. The development of high efficiency OER catalysts that consist of stable, inexpensive and earth abundant elements is of highest interest and presents an ongoing global research challenge. Heretofore, various different metal-based oxides/ (oxy) hydroxides, chalcogenides and phosphides as well as organometallic compounds have been widely explored toward water oxidation reaction [34,35,36,37,38,39,40,41, 42, 43, 44, 45]. Among them, transition metal phosphides such as $Ni_2P$ [38, 46, 47], $Co_2P$ [48, 49], and CoP [50], and ternary metal phosphides (NiCoP, CoMnP and $NiMoP_2$) [26, 51, 52, 53] emerged as a new class of highly active OER catalysts [36,41, 54]. Iron based phosphides, however, have rarely been used for OER electrocatalysis. Iron, substantially cheaper and more abundant than Ni and Co, is considered to be the most economical element among all transition metals.

Recently, commercial stainless steels, composed of *e.g.* Fe, Ni, Cr, Mn and Co emerged to be attractive OER catalyst candidates[13]. For example, Chatenet and co-workers [55] found that the OER activity of commercially available 316L stainless steel after long-term usage as oxygen evolving electrode in aqueous lithium-air batteries increased substantially. To the best of our knowledge some of our group were the first who tailored steel intentionally into outstanding efficient and stable OER electrocatalysts [56, 57, 58] based inter alia on thin Ni, Fe oxide layers firmly attached to the iron-based matrix. Benchmark OER properties were shown at pH 13 and pH 14 as the overpotential amounted to 212 mV at 12 mA cm$^{-2}$ in 1 M KOH, η= 269.2 mV at 10 mA cm$^{-2}$ in 0.1 M KOH, respectively. Very recently, hydrothermal/electrochemical oxidation of Cr Ni based steel[59] as well as the surface oxidation of Ni42[60] and X20CoCrWMo10-9//$Co_3O_4$ steel [57, 61] also proved to be suitable to achieve acceptable OER catalysts efficiently, and stable and working over a wide pH range. Besides surface oxidation, phosphorization and sulfurization have been applied to steel i.e. Steel 316 (basically consisting of Ni, Cr and Fe) in order to gain better overall OER characteristics[62]. All these researches indicate that stainless steels have great potential for application in water splitting. We chose standard Carbon-Manganese steel (S235) which consists of ~ 98 % Fe and 1.5 % Mn as starting material to synthesize costless Fe based phosphides. To the best of our knowledge, it can be considered as the cheapest Fe source in general and can be directly used as a working electrode negating the necessity of a polymer binder (*e.g.* Nafion). The corresponding surface oxidation modification has been reported by our group[63]. Oxidized mild S235 steel showed enhanced OER performances in comparison with untreated steel (347 mV overpotential at 2 mA $cm^{-2}$ current density in 0.1 M KOH electrolyte) with electrocatalytic properties improved after surface oxidation.

The results presented here not only verify that the phosphorized S235 steel exhibits excellent catalytic OER properties but also demonstrates outstanding long-term durability not only under alkaline but also under neutral conditions thus pointing out the great potential application of modified steel S235 in the development of new energy applications.

**Results and discussion**

After removing the surficial layer of S235 steel by rubbing with 600 SiC sanding paper, the polished steel was phosphorized under the protection of Ar flow, as schemed in Figure 1a. Along with the phosphorization treatment, the colour of the polished steel S235 has changed from grey to black (see inset, Figure 1b). In order to investigate the transformation during the phosphorization treatment, X-ray powder diffraction (XRD) measurements were conducted on the phosphorized steel plates and on untreated S235 steel (Figure 1b). 3D S235-P steel exhibits three strong and narrow diffraction peaks at 2-theta 40.3°, 44.6° and 64.9°, which can be well indexed to the (111) plane of $Fe_2P$ (JCPDS No. 85-1725), as well as to the (110) and (200) planes of metallic iron (JCPDS No. 05–0696). The weaker diffraction peaks at 2-theta 41.2°, 51.5° and 52.8°, 42.8° and 45.8° are in good agreement with the (201), (002) and (300) planes of $Fe_2P$ (JCPDS No. 85-1725) and (112) and (411) planes of $Fe_3P$ (JCPDS No. 89-2712) [64, 65, 66]. In contrast, untreated S235 steel shows two distinct XRD peaks (Figure 1b) at 2-theta 44.6° and 64.9° belonging to (110) and (220) planes of Fe (JCPDS No. 05–0696). As expected signals that can be assigned to Mn are missing which can be explained with the low amount of manganese in S235 steel.

The morphology and microstructure of the obtained 3D S235-P steel were characterized using SEM and TEM. As shown in Figures 2a, 2b, the surface of 3D S235-P steel consists of a well-defined 3D porous network structure with several hundred nanometer of the interconnected pores, enabling diffusion of the electrolyte into the pores as well as an effective contact with the catalysts underneath. This morphological transformation can likely be attributed to the structural collapse and reconsolidation of temporary formed nano- and micro-scaled particles during the reaction of S235 steel with $PH_3$, a decomposition product of $NaH_2PO_2$. The corresponding EDS elemental mapping images in Figures 2c−2e reveal the successful and homogeneous phosphorization of untreated S235 steel. The surface of the modified steel was found to be quite smooth as derived from high resolution SEM (Figure 2b). A cross-sectional analysis of the 3D S235-P steel exhibited a catalyst layer around 40 μm in thickness (Figure S1). The detailed TEM investigation was carried out to examine the microstructure of the phosphorized steel. Figure 2f shows a low magnification TEM image of an individual sheet of the phosphorized steel, which exhibits an analogous rectangle shape and consists of stacking faults derived from the different contrast at the edges. We selected an area at the central part (the marked red rectangle area) for the electron diffraction (SAED) analysis, high-resolution TEM (HR-TEM) and energy-dispersive X-ray spectroscopy (EDX) spectrum analysis. From Figure 2g, it is evident that the prepared 3D S235-P steel is highly polycrystalline and contains many dislocations and stacking layers. The synthesized $Fe_2P$ and $Fe_3P$ was identified from the HRTEM image (Figure 2h). The lattice fringes have an inter-plane distance of 0.294 nm and 0.167, which was indexed to the (110) planes and (211) planes of $Fe_2P$ lattice. The calculated inter-plane distance of 0.210 nm pertains to the (112) planes of $Fe_3P$ lattice. More importantly, the intense connection between $Fe_2P$ and $Fe_3P$ will play a critical role in the electron transport during the OER process. In addition, the corresponding elemental mapping EDX spectrum (Figure 2i) shows that the elements of Fe, P and Cu exist in the steel sample. Here, the element Cu may stem from the TEM copper grid.

The phorphorization resulting in sample 3D S235-P steel was further confirmed by X-ray photoelectron spectroscopy (XPS). Quantitative XPS analysis shows a substantial amount of phosphorous after phorphorization: 48.9% of P, 51.0% of Fe and 0.1% of Mn. Figure 3 displays high resolution P 2p, Fe 2p, and Mn 2p spectra of sample 3D S235-P steel. The P 2p core level spectrum in (Figure 3a) comprises of two 2p peaks at ~129.6 eV (P-Fe bonds) and ~134 eV (P-O bonds), as marked by the grey vertical bars. The top panel of Figure 3 presents the P 2p core level spectrum, and the 2p positions of P-Fe bonds (~129.6 eV), and P-O bonds (~134 eV) are also shown by gray vertical bars[67,68,69]. The P 2p spectrum of sample 3D S235-P before OER comprises of two peaks, indicating the presence of Fe-P bonds and a strong main peak at 134 eV which can be separated into two peaks (c.f. Figure 3a), therefore confirming the presence of at least two higher P oxidation states. This outcome is also supported by the Fe 2p spectrum (Figure 3b) showing a small peak at 706.8 eV corresponding to low valence Fe-P bonds [69], followed by a peak of maximum intensity at around 710 eV typical for Fe3+ ions. This finding is also supported by the Fe 2p spectrum (Figure 3b), here we find a small peak at 706.8 eV corresponding to low valence Fe-P bonds [69], followed by a peak of maximum intensity at around 711 eV, which again can be separated into tow peaks located around 710 eV and 712 eV. Those peaks may be assigned to $Fe^{2+}$ and $Fe^{3+}$ ions that are likely to stem from a mixture of iron oxides and iron hydroxides, respectively. The absence of intense charge transfer satellites (which may be de-convoluted by fitting (c.f. Figure 3b)) similar to the Fe 2p core level spectrum of magnetite[70] indicates a mixed $Fe^{2+}/Fe^{3+}$ state. As to Mn we find small amounts of oxidized species at the surface of sample 3D S235-P, albeit the exact valence state cannot be determined unambiguously (Figure 3c).

The XPS experiments verify the successful allocation of Fe, Mn and P on the unique 3D porous structure. Therefore, SEM, XRD, TEM and XPS results all confirm the formation of a novel self-standing 3D porous hybrid-material (3D S235-P steel) unmasking the phorphorization as an efficient way to modify the stainless steel. We claim that these microstructural characteristics are responsible for the improvement of OER properties described in detail in the next paragraphs when compared to the OER properties of untreated S235 steel S235.

The electrocatalytic activities of the phosphorized steels towards OER in 0.1 M KOH were evaluated by using a three-electrode apparatus in alkaline solution. As shown in Figure 4a, the cyclic voltammetry (CV) curves reveal that the 3D S235-P steel exhibits much higher current density at given potential and lower onset potential than S235-P-450, S235-P-650 and the untreated S235 steel do. The OER catalytic current density of 3D S235-P steel increases significantly above 1.47 V *vs.* RHE in Figure 4a (the red curve). The overpotential required to deliver a current density of 10 mA cm$^{-2}$ is η = 295 mV, which is lower than that of the S235-P-650 and S235-P-450 steel (η = 318 mV and 441 mV, respectively), indicating that the phosphorized steel prepared at 550 ℃ is much more suitable for supporting OER. In contrast, the negligible anodic current density derived from the CV curve (for potentials up to 1.7 V *vs.* RHE), shows that untreated S235 steel is not catalytically active. Broad and very weak oxidation peaks were found between 1.3~1.45 V vs RHE of 3D S235-P steel and 1.45~1.6 V vs RHE of S235-P 450 steel, which might be attributed to the transitions from Fe$^{II}$ to Fe$^{III}$, and the oxidation of Fe to Fe$^{II}$ because of the partially phosphorized stainless steel [71]. Generally, oxygen evolution is known to take place solely on metal-oxides and not on metals. However, we believe that temporarily occurring charge effects may also have caused this weak increase of the current density. This effect is known to play a role when the layer is thick as it is indeed the case (Figure S1). Previously, our group reported on the OER properties of S235 steel oxidized by chlorine gas[63], where oxidized S235 steel in comparison with untreated S-235 steel exhibited slightly enhanced OER properties with η = 347 mV at 2 mA cm$^{-2}$ in 0.1 M KOH electrolyte.

Figure 4b displays Tafel plots of the steel samples. Throughout the potential range the 3D S235-P steel exhibits a much stronger current/voltage behaviour than S235-P-450, S235-P-650, and untreated S235. A substantial horizontal shift of the Tafel lines of the phosphorized steels Figure 4b (red, blue and green line) relative to the one of untreated S235 steel (black line) underpins the substantial improvement of the electrocatalytic performances of steel S235 upon phosphorization. The difference between the two samples (3D S235-P/S235-P650 and S235-P450/S235) in the required overpotential increases substantially with increasing current density, the curves move increasingly apart from each other towards the high-potential region. Compared with untreated S235 steel, for example, 3D S235-P steel shows a difference of 133 mV at 2 mA cm$^{-2}$, increasing to 350 and 463 mV at 8.5 and 16.3 mA cm$^{-2}$. We attribute this increase to the differences of the surface composition of the specimens. Interestingly, the 3D S235-P steel exhibits a Tafel slope of 68.7 mV dec$^{-1}$, which is much lower than the S235-P-650 steel (88.4 mV dec$^{-1}$), S235-P-450 steel (135.5 mV dec$^{-1}$) and 198.7 mV dec$^{-1}$ of the untreated S235 steel.

This suggests favourable OER kinetics for the 3D S235-P steel. After IR correction, the corresponding Tafel slope shift to 67.3, 76.8, 123.9 and 196.6 mV dec$^{-1}$ (Figure S2) and all steel samples exhibit dual Tafel behaviour. This might be the result of the unique 3D steel structure that enhances the OER properties substantially through generating a large electrochemically active surface area and numerous active sites consistent with the electron microscope measurements. Sample 3D S235-P steel (prepared at 550 ℃) presents the best overall OER properties consequently the long-term performance tests were solely applied to 3D S235-P steel specimen.

The long-term electrochemical durability of 3D S235-P steel, a vital feature to evaluate the activity of catalysts in practical applications, was performed in 0.1 M KOH electrolyte at a constant current density of 10 mA cm$^{-2}$. As displayed in Figure 4c, the potential of 3D S235-P steel exhibits a slight potential increase from 1.560 V to 1.575 V through the first 2000 s of chronopotentiometry, and then falls back to about 1.560 V in the flowing 2000 s. This is likely ascribed to the increased electrocatalytic active species enabling in alkaline solutions. It suggests that an activation process could occur in the initial 4000 s of stability measurements converting iron phosphide to OER active Fe oxide species. This hypothesis was strengthened by XPS spectroscopy (Figure S3a-c; grey curves) carried out after the OER polarization measurements (see Section OER mechanism). The 3D S235-P steel (the red line in Figure 4c) maintains a sufficient current-voltage stability (about 1.556 V at 10 mA/cm$^2$ with average η = 326 mV) throughout the chronopotentiometry measurement, which suggests that the unique 3D porous structure contributes to the ease of oxygen release. The OER-based overpotential of 3D S235-P steel ($η_{10}$ = 326 mV) is lower than those of other Fe based catalysts recently reported, such as FeP nanorods dispersed on carbon fibre paper ($η_{10}$ = 350 mV)[69], electrodeposited amorphous FeOOH ($η_{10}$ > 420 mV)[72], Ni doped FeOOH thin films ($η_{10}$ > 340 mV)[73], sea-urchin-like (Co$_{0.54}$Fe$_{0.46}$)$_2$P ($η_{10}$ = 370 mV)[74], Fe$_{2-x}$Mn$_x$P nanorods (($η_{10}$ > 480 mV) [75], NiFeO$_x$ film ($η_{10}$ > 350 mV) [33], and exfoliated nanosheets of NiFe LDHs ($η_{10}$ > 350 mV) [76] (Table S1, ESI†). The OER stability can also be derived from a comparison of polarization curves before and after 1000 continuous CV scans within the potential range 1.50–1.65 V (*vs.* RHE) (Figure 4d). Almost no drop of activity can be found after 1000 cyclic sweeps, confirming the robustness of 3D S235-P steel for extended OER catalysis. Therefore, the self-standing 3D S235-P steel is intrinsically stable and gives more possibility for the application in water electrolysis industry, highlighting the utility value of this kind of inexpensive modified steels.

To further investigate the OER kinetics activity, the electron transfer resistance was conducted by electrochemical impedance spectroscopy (EIS) measurements at an offset potential that ensures oxygen evolution (1.55 V *vs.* RHE, Figure 5). A circuit consisting of a resistor in series with a parallel combination of a resistor and a constant phase element (inset of Figure 5) was used for all EIS experiments. The smaller the circular arc is in Nyquist plot, the lower the electron transfer resistance will be. Typically, the real axis value at high frequency intercept can be interpreted as the series resistance ($R_s$) of the electrolyte, the real axis value at low frequency intercept can be assigned to the sum of the electrolyte resistance and charge transfer resistance ($R_{ct}$) of the redox reaction. The Nyquist plots (Figure 5) show that the $R_s$ value decreased from 6.2 to 4.5 Ω after the S235 steel was phosphorized, indicating the improved conductivity after phosphorization treatment. The charge transfer resistance of 3D S235-P steel is 1.9 Ω, which is much lower than the one of untreated S235 steel ($R_{ct}$ = 8.4 Ω) suggesting that 3D S235-P steel has a lower resistance and superior electron-transfer kinetics for OER, thereby resulting in the faster charge collection and transport for ensuring enhanced catalytic properties. More importantly, determination of the electrochemical active surface area (ECSA), a critical parameter that represents the amount of active sites in electrocatalysts, was employed to evaluate the electrochemical activities of 3D S235-P steel using a cyclic voltammetry (CV) technique. As shown in Figure S4, the double-layer capacitance ($C_{dl}$) of 3D S235-P steel is about 46.1 mF cm$^{-2}$, which is much higher than what can be determined for the untreated S235 steel (about 0.2 mF cm$^{-2}$). These differences could be attributed to the phoshporization treatment that introduces 3D interconnected nanoporous surfaces with a large electrochemically active surface area and numerous active sites, thereby facilitating faster charge collection and transport thus ensuring excellent catalytic properties.

In order to confirm whether the oxidation currents, determined during the electrochemical measurements are exclusively associated with water oxidation, it is essential to quantify the real oxygen evolution efficiency upon evaluation of the charge to oxygen conversion rate. The Faradic efficiency of 3D S235-P steel was conducted by direct fluorescence-based detection of the concentration of evolved oxygen during (duration: 2000 seconds s) chronopotentiometry measurements at constant current density of 5 mA cm$^{-2}$, and 10 mA cm$^{-2}$ in 0.1 M KOH solution.

The plot of the dissolved oxygen (mg/L) as a function of time (s) (black square dot) shows a good agreement with the possible theoretical increase of dissolved oxygen on the basis of 100% charge to oxygen conversion (100% Faradaic efficiency; see red line in Figure 6a). The faradic efficiency was found to be about 82.1% after 2000 second s running time with a very stable overpotential at 323 mV (Figure 6b), therefore attesting excellent electrocatalytic oxygen evolution activities of 3D S235-P steel. A significant higher Faradaic efficiency (89.6%) close to the theoretical possible value of 100% was achieved (2000 s of chronopotentiometry) at 5 mA cm$^{-2}$ (Figure 6c), using 265 mV overpotential (Figure 6d). There is a deviation of the Faradaic efficiency of 3D S235-P steel measured at 10 mA cm$^{-2}$ and 5 mA cm$^{-2}$. This phenomenon (a decreasing Faradic efficiency at higher current density), also reported by other groups[77], might be caused by a slight measurement inaccuracy: the oxygen cannot be detected as fast as it is formed. As reported by Qiu et al., Faradaic efficiency of OER for Ni–Fe containing nanoparticles in 1 M KOH decreased from 97% at 1 mA cm$^{-2}$ to 43% at 10 mA cm$^{-2}$. In addition, OER Faradaic efficiency of untreated S235 steel, as previously reported by our group[63], at 2 mA cm$^{-2}$ was 67% after chronopotentiometry measurement in 0.1 M KOH, which is much lower than the 82.1% of phosphorized 3D S235-P steel.

OER electrocatalysts that exhibit enhanced properties in neural media are also very important and attractive, for example potentially allowing the future development of applications with direct usage of seawater as electrolyte. Hence, OER performances of 3D S235-P steel in 0.1 M KH$_2$PO$_4$/K$_2$HPO$_4$ at pH 7 were conducted as displayed in Figure 7. The current density of 3D S235-P steel (Figure 7a) increases to 15.3 mA cm$^{-2}$ at a potential of 1.9 V *vs.* RHE with an overpotential of 308 mV at 2 mA cm$^{-2}$, which is much lower than untreated steel S235, as well as that of surface oxidized S235 (with a overpotential of 462 mV at 1 mA cm$^{-2}$ in pH 7)[63] exhibited. The 3D S235-P steel not only shows improved current density, but also exhibits superior durability with respect to chronoamperometry experiments (Figure 7b). The current density maintained at a relative stable value without much increment through 40000 s of chronopetentiometry was carried out at 2 mA cm$^{-2}$ in pH 7 conditions. To confirm these results, we examined the mass loss of the electrode occurring through a long-term chronopotentiometry experiment (40000 s; j= 2 mA/cm$^2$) and, in addition conducted an ICP-OES analysis of the pH 7 electrolyte that was used for this long term OER testing (40000 s). Three test runs were evaluated showing an average mass loss of 0,10 mg (total electrode area: 2 cm$^2$), and the average concentration of Mn and Fe in solution was close to 0 mg/mL (Table S2). Thus, the 3D S235-P steel works as a high efficiency electrocatalyst both in alkaline and neutral conditions.

3D S235-P steel samples were further investigated using XRD, SEM, contact angle, XPS, Raman and FTIR analyses after 2000 s of usage in electrochemical testing to disclose the OER mechanisms. The $Fe^{3+}$ oxidation state (Figure S3a) including the prominent charge transfer satellites can be derived from the Fe-2p XPS spectrum shown in Figure S3. There is no peak at 706.8 eV, and hence no Fe-P bonds are present, which can also be seen in the corresponding P 2p core level spectrum (Figure S3b). It is therefore reasonable to assume that $M^{3+}$ species such as $M_2O_3$ and M(O)OH (M = Fe, Mn), acting as active species, have been formed during the usage as water oxidation catalyst. The unique 3D porous frame structure was retained very well after these electrocatalytic measurements (Figure S4a Figure S5a). However, instead of the smooth surface (Figure 2b), high-resolution SEM image (Figure S4a Figure S5a) reveals a layer of rough, edged species emerged in the high-resolution SEM image, which could also be detected though the XRD results (Figure S4d Figure S5a). The peak that corresponds to $Fe_2P$ is significantly reduced compared with the catalyst prior to the polarization experiments, suggesting that new active iron-based species have been generated during OER process. In addition, the corresponding elemental mapping images (Figure S4d Figure S5a) disclose the uniform distribution of Fe, Mn, P, and O, suggesting that further oxidation occurred during catalysis. Contact conditions between catalyst and electrolyte are a critical factor towards the catalytic performance. As shown in Figure 8 (a, b), the contact angle decreases significantly after phoshporized treatment, demonstrating that the 3D S235-P steel shows better wettability than untreated S235 steel does. This can be attributed to the unique 3D porous structure allowing better accessibility of active sites and facilitates the uptake of oxygen adsorbate, resulting in significantly greater water adsorption and water oxidation.

*The mechanism of the formation of the active surface layer*

The FTIR experiments performed with sample 3D S235-P steel are in good agreement with these findings. The FTIR absorption peaks (Figure 8c) at 569, 717, and 792 $cm^{-1}$ can be attributed to the stretching vibration of metal-oxygen (Fe–O) bond, Fe–OH mode, and Fe–O–H bending vibrations of α-FeOOH [78,79,80,] respectively, indicating that Fe is surrounded by oxygen. In addition, the broad absorption band at 3403 $cm^{-1}$ was assigned to the O–H stretching vibration of hydroxide species and physically adsorbed water. Moreover, the absorption peaks at around 1347 and 1498, 1633 cm$^{-1}$ may be caused by O–H and P–O bending vibrations of Fe containing species[81, 82]. Furthermore, after the electrochemical measurement a Raman investigation, a reliable method to distinguish phases of materials that do not have a long-range order, was conducted. As shown in Figure S5, peaks situated at 262 and 509 cm$^{-1}$, and the strong bands at 715 cm$^{-1}$ and 377 cm$^{-1}$ can be indexed to α-FeOOH, β-FeOOH and γ-FeOOH[83, 84] the weak bands located at 301, 530 and 661 cm$^{-1}$, and 388 cm$^{-1}$ were caused by $Fe_3O_4$ and α-$Fe_2O_3$ [85, 86] only a small peak at 488 cm$^{-1}$ pertaining to $Fe_3P$ [87] can be found, which confirm the formation and coexistence of $M_2O_3$, $M_3O_4$ and M(O)OH (M = Fe, Mn) during OER process. Because of the relatively low amount of Mn, no signal could be detected. These results verify that oxidation occurred during electrocatalysis that coincides with a depletion of the P-consisting layer and ends up in a Fe-oxide based OER catalyst. Notably: A similar strategy for the generation of an active OER layer was reported by Lee et al. very recently [88]. A chalcogenide (Fe–S) overlayer generated by sulfurization on the SS surface was found to play a critical role as a precursor layer in the formation of an active surface during water oxidation. We discussed the mechanism of OER on oxidized steel S235 in our recent report [63] and based on the fact that comparable active species can be found on phosphorized S235 steel after usage as an OER electrocatalyst the basic mechanism should not differ from the earlier suggested one. However, we assume that the surface concentration of real active species is increased due to the pre-phosphorization treatment as discussed above.

In conclusion, to develop high efficiency cost-effective catalysts based on earth abundant elements we established a method to produce self-sustaining porous 3D S235-P nanostructures by phosphorizing standard S235 steel at relative low temperature. Under alkaline conditions, the phosphorized 3D S235-P steel exhibits high OER performance with high current density, excellent catalytic activities and superior stability. These features are enhanced by the 3D interconnected nanostructure and yield an excellent wettability. The 3D S235-P steel also shows improved OER performances in neutral pH solution compared with untreated S235 steel. This means that the 3D S235-P steel plate can act as a promising catalyst candidate for electrocatalytic water splitting, highlighting that a one-step generation route can be efficient to develop earth abundant elements based OER electrodes. We believe this method will help to pave the way for utilizing cost-efficient, commonly available elements into efficient catalysts to produce energy, therefore potentially helping to alleviate cost derived bottlenecks in this timely area of energy research.

## Experimental Materials and Methods

**Materials and Methods.** Standard Carbon-Manganese steel S235 plates were purchased from WST Werkzeug Stahl Center GmbH & Co. KG, D-90587 Veitsbronn-Siegelsdorf, Germany, with a uniform size of 35 mm long, 9 mm wide and 1mm thick. Prior to surface modification, the steel plates were polished with grit 600 SiC standing paper, washed with acetone, ethanol and deionized water upon sonication for 15 min, and were allowed to dry for 50 min at room temperature. They were stored in vacuum for further use. For the phosphorization, a piece of clean S235 steel plate was positioned at the centre heating zone of the tube furnace and 2 g $NaH_2PO_2·H_2O$ (ACROS ORGANICS, Belgium) was put at an appropriate upstream side. Subsequently the tube was flushed with Ar for 30 min, heated up to 550 ℃ at a heating rate of 5 ℃ $min^{-1}$, and maintained at that temperature for 2 h. After completing reaction and cooling down to room temperature, the phosphorized self-standing 3D S235-P steel plate was taken out of tube. The loading mass of phosphorized layer was determined to be 0.68 mg $cm^{-2}$ using a high precision microbalance. For the control experiments, samples prepared at 650 and 450 ℃ were also obtained and noted as S235-P-650 and S235-P-450, respectively.

**X-ray diffraction (XRD)** patterns were recorded on a PANalytical X'Pert Pro MRD diffractometer, which is specially used for analysing thin films grown on substrate, equipped with an Eulerian cradle in reflection mode, operating with CuKa radiation at 40 kV and 40 mA. SEM investigation was performed by using a Zeiss Auriga scanning electron microscope with INCA 350 energy-dispersive X-ray spectroscopy (EDS) (Oxford Instruments).

**Transmission electron microscopy (TEM)** and high-resolution transmission electron microscopy (HR-TEM) were carried out on a probe-corrected transmission electron microscope operating at 200 kV (JEOL JEM-2100, Gatan CCD-camera). The TEM sample was prepared by sonicating the phosphorized steel plate in water for 10 s and dropping the resulted solution on top of the copper grid.

**X-ray photoemission spectroscopy (XPS)** was performed at room temperature on a Phoibos HSA 150 hemispherical analyser equipped with a standard Al Ka source with a 0.4 eV full width at half-maximum workstation. The spectra were calibrated using the carbon 1s line of adsorbed carbon (EB = 285.0 eV). The contact angle was performed with DROP SHAPE ANALYSIS SYSTEM DAS 10 Mk2 (KRÜSS). The spectra were calibrated using the carbon 1s line of adsorbed carbon ($E_B$ = 285.0 eV).

**Infrared Spectroscopy (IR) spectra** were recorded on a Bruker Vertex 70 equipped with the ATR system, Golden Gate. For this measurement, the phosphorized S235 steel plate was ground by grit 800 sanding paper till the black outer part was completely removed from the steel substrate. Raman spectra were carried out with a Raman microscope WITec Alpha 300R (30 cm focal length and 600 grooves per mm grating spectrometer) equipped with an EM-CCD (Andor Newton DU970N–BV-353) under 632.8 nm line of a He-Ne laser with a power of 1.5 mW. Powder sample were removed from the contaminated sanding paper by ultrasonication in distilled water.

**ICP-OES analysis** The pH 7 electrolyte solution samples were analysed by inductively coupled plasma optical emission spectrometry (ICP-OES, iCAP™ 7400 Duo equipped with MiraMist® Teflon nebulizer, Thermo Fisher Scientific Germany BV & Co KG) according DIN EN ISO 11885:2009-09. Concentrations of selected elements were determined at wavelengths of 259.9 nm (Fe) and 257.6 nm (Mn). The calibration standards were prepared according the matrix of the analyte solution and contained 0.001, 0.01, 1.0 and 10 mg L-1 of the selected elements.

**Electrochemical Measurements.** All electrochemical measurements were carried out in a standard three-electrode system using a Potentiostat Interface 1000 from Gamry Instruments (Warminster, PA, USA). As-prepared 3D S235-P steel plate was directly used as a working electrode (WE) on which a precise surface area of ~2 cm$^2$ was defined by insulating tape (Kapton tape). A platinum wire electrode (4 × 3 cm geometric area) was employed as the counter electrode (CE) and a reversible hydrogen reference electrode (RHE, HydroFlex, Gaskatel Gesellschaft für Gassysteme durch Katalyse und ElektrochemiembH. D-34127 Kassel, Germany) was utilized as the reference standard electrode (RE). And all voltages measured were quoted against this reference electrode. The reference electrode was placed between working electrode and counter electrode with a distance of 2 mm and 5 mm to each other. All measurements were performed at room temperature. Cyclic voltammetry (CV) polarization curves were recorded in 90 mL electrolyte in a 100 mL glass beaker under stirring (500 rmp) in pH 13 (0.1 M KOH) and pH 7 (0.1 M $KH_2PO_4/K_2HPO_4$), respectively. The corresponding potential region was 1.0-1.7 V *vs.* RHE at pH 13 and 1.0-1.9 V *vs.* RHE at pH 7. The scan rate was 20 mV s$^{-1}$ and the step size was set to 2 mV. The chronopotentiometry measurements were conducted at a constant current density of 10 mA cm$^{-2}$ in 800 mL of 0.1 M KOH in a 1000 mL glass beaker, and at 1 mA cm$^{-2}$ in 500 mL of 0.1 M 0.1 M $KH_2PO4/K_2HPO_4$ in a 800 mL glass beaker. The durability analysis was taken by CV scanning from 1.50 to 1.65 V (*vs.* RHE) at 20 mV s$^{-1}$ for 1000 cycles. Tafel plots were derived from the average voltage values of 200 s chronopotentiometry scans at current densities of 0.80, 1.46, 2.03, 2.66, 4, 5.80, 7.07, 8.5, 10, 12.93, 16.3 and 20 mA cm$^{-2}$ for measurements at pH 13 and pH 7. Electrochemical impedance spectra (EIS) of samples were measured at 1.55 V (*vs.* RHE) in the frequency range of 0.1-50 000 Hz in 0.1 M KOH electrolyte with an Autolab PGStat 20 potentiostat, controlled by FRA Windows software (Frequency Response Analysis for Windows version *4.9.007*). In order to accurately investigate the Tafel behavior, we corrected Ohmic losses manually by subtracting the Ohmic voltage drop from the measured potentials in Figure 4b, according to Ohm's Law and series resistance (Rs). The Rs was derived from the EIS Nyquist plot since the first intercept of the main arc with the real axis in Figure 5. The IR-corrected potentials are denoted as E-IR (Figure S2). The electrochemically active surface area was conducted by electrochemical double layer capacitances ($C_{dl}$) measurements. A potential range from 0.565 to 0.665 V vs RHE where no faradaic process occurs was selected for the capacitance measurements at different scan rate (20, 40, 60, 80, 100 and 120 mV s$^{-1}$). Then, the capacitive currents ($\Delta J_{|Ja-Jc|}$)/2 were measured at 0.615 V vs RHE and plotted as a function of scan rate, these disperse dots were then linearly fitted to a line, the obtained slop was the geometric $C_{dl}$.

**Faradaic efficiency.** The measurement of Faradaic efficiency was carried out by detecting the oxygen concentration in electrolyte depending on the time during chronopotenmetry at constant current in alkaline solution under stirring (300 rmp) in accordance with the procedure reported by Schäfer *et al.*[35]. The oxygen concentration was recorded via so-called fluorescence quenching method with an optical dissolved oxygen (OD) sensor (Multi 3420 IDS from WTW, Weilheim, Germany) interfaced to a personal computer. The Faradaic efficiency of 3D S235-P steel were performed at constant current densities of 10 mA cm$^{-2}$ and at 5 mA cm$^{-2}$ in 0.1 M KOH in a four necked 2300 mL glass vessel with WE, RE, CE and OD inserted. Before measurement, the vessel was filled with alkaline electrolyte, and continuously purged with Argon for 3 hours at a constant flow rate of 0.3 cm$^3$ s$^{-1}$ till the dissolved oxygen was as low as 0.06, and 0.10 mg L$^{-1}$, respectively. The whole system was completely sealed with glass stoppers before starting the measurement. The dissolved oxygen concentration and electrochemical data were recorded upon device (Multi 3420 IDS from WTW) and Potentiostat Interface 1000 from Gamry Instruments (Warminster, PA, USA) simultaneously. The theoretical 100% Faradaic efficiency was calculated with a line equation: y = 7.239 × 10$^{-4}$ x + 0.06 (y = Dissolved oxygen (mg L$^{-1}$); x = time (s)) at 10 mA cm$^{-2}$ and y = 3.688 × 10$^{-4}$ x + 0.10 (y = Dissolved oxygen (mg L$^{-1}$); x = time (s)) at 5 mA cm$^{-2}$, respectively, as shown red line in Figure 2b and 2d.

## Acknowledgments

This work was supported by the European Research Council (ERC-CoG-2014; project 646742 INCANA) and the German Research Foundation (INST 190/164-1 FUGG).

## Supporting Information Available

The details thickness of the average phosphorized layer and electrochemical active surface area (ECSA) SEM, XRD, Raman and XPS results of the long-term measured 3D S235-P steel are provided in Supporting Information.

**Figure captions**

Figure 1. (a) Scheme of the synthesis process of 3D S235-P steel; (b) Typical XRD patterns of the as-prepared 3D S235-P steel and untreated S235 steel with the digital photo of phosphorized S235 steel inserted.

Figure 2. SEM (a, b) of porous 3D S235-P steel and the corresponding elemental mapping images (c, d, e); typical TEM image (c) (f) and the SAED (d) (g), the HR-TEM image (h) and the corresponding elemental mapping EDS (i) of 3D S235-P steel (red marked square area).

Figure 3. High-resolution XPS patterns for 3D S235-P steel for P 2p (a), Fe 2p (b) and Mn 2p (c).

Figure 4. Electrocatalytic properties of the as-prepared 3D S235-P steel for oxygen evolution in alkaline electrolyte (0.1 M KOH) in comparison with S235-P-650 steel, S235-P-450 steel, and untreated S235 steel: (a) Cyclic voltammetric plots at scan rate of 20 mV s$^{-1}$ and 2 mV step size. (b) Tafel plots based on 200 second chronopotentiometry scans at current densities 0.80, 1.48, 2.03, 2.67, 4.12, 4.98, 5.82, 7.07, 8.50, 9.98, 12.93, 16.36, and 20.34 mA cm$^{-2}$. (c) Long term chronopotentiometric measurement of 3D S235-P and untreated S235 steel performed at a current density of 10 mA cm$^{-2}$. (d) Long-term cycling tests of the 3D S235-P steel.

Figure 5. EIS spectrum spectra of 3D S235-P steel and untreated S235 at potential 1.55 V (*vs.* RHE) in 0.1 M KOH.

Figure 6. (a, c) Faradaic efficiency of the 3D S235-P steel performed in 0.1 M KOH electrolyte for 2000 s at a constant current density of 5 and 10 mA cm$^{-2}$, respectively; (b, d) the corresponding chronopotentiometric measurements were recorded.

Figure 7. Electrocatalytic properties of 3D S235-P steel for oxygen evolution in neutral electrolyte (0.1 KH$_2$PO$_4$/K$_2$HPO$_4$) in comparison with the untreated S235 steel: (a) Cyclic voltammetric plots at scan rate of 20 mV s$^{-1}$ and 2 mV step size. (b) Long term chronopotentiometric measurements of 3D S235-P steel performed at a current density of 2 mA

cm$^{-2}$.

Figure 8. Contact angle images of untreated S235 steel (a) and 3D S235-P steel (b), and FTIR (c) of 3D S235-P steel before and after long-term chronopotentiometric measurement.

Figure 1

(a)

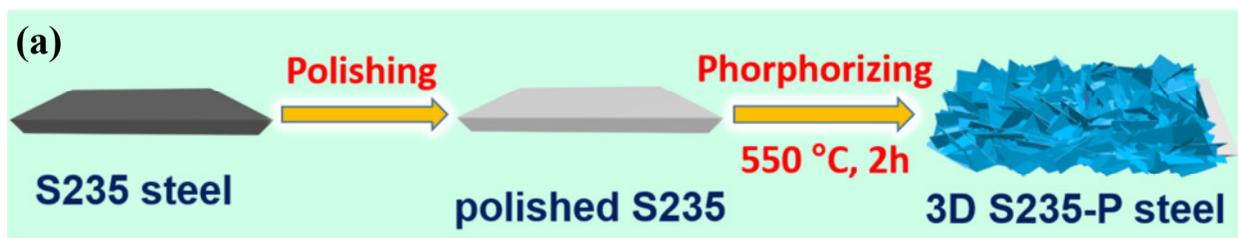

(b)

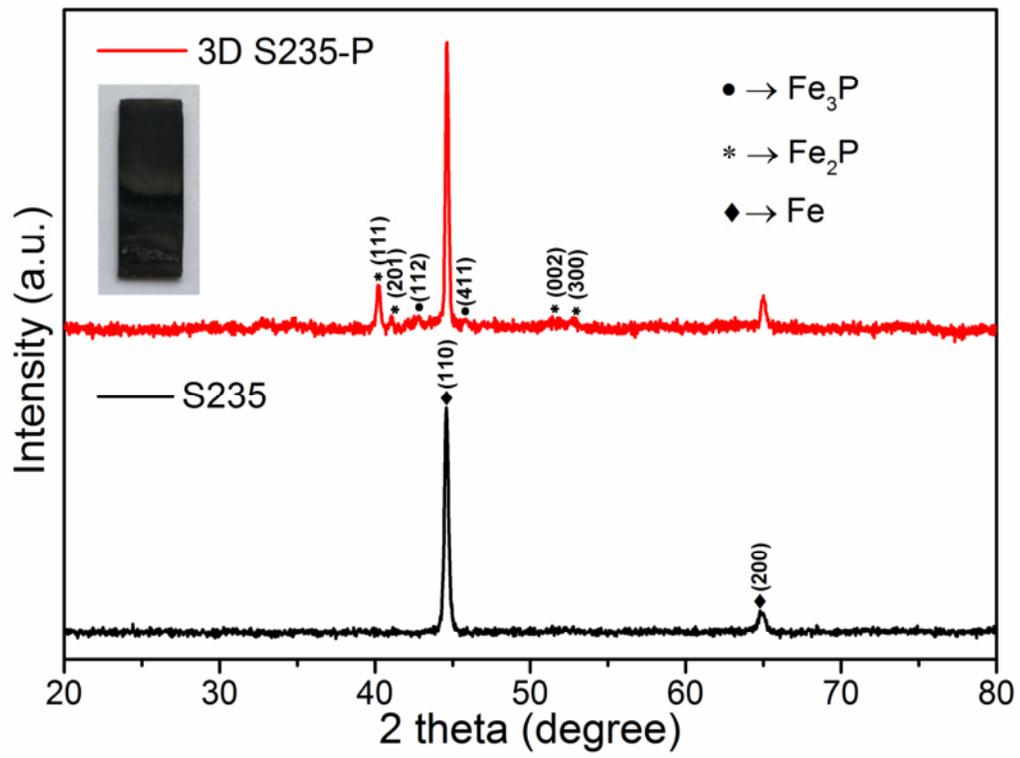

Figure 2

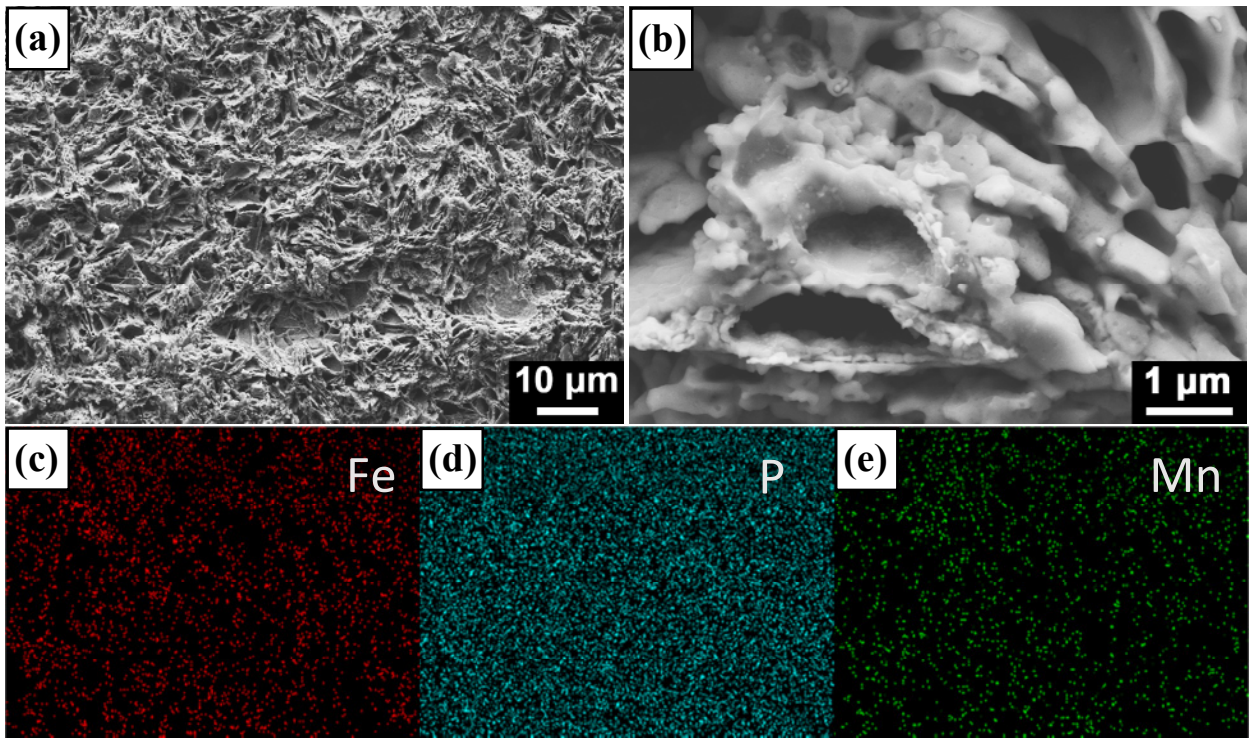

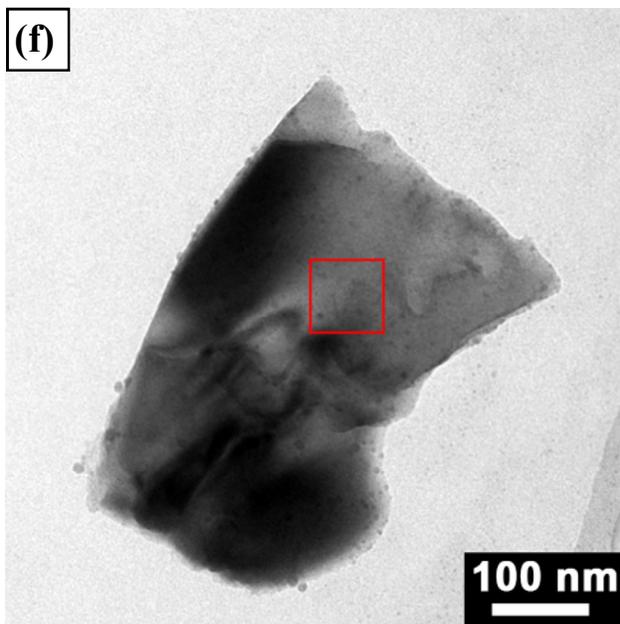
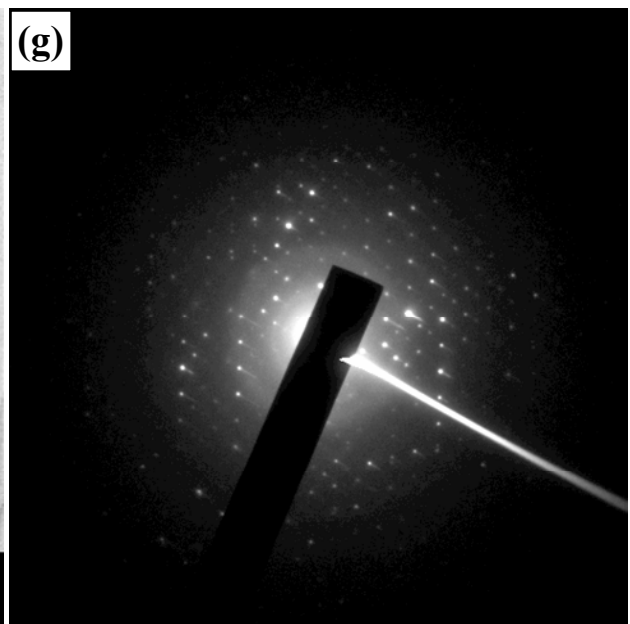
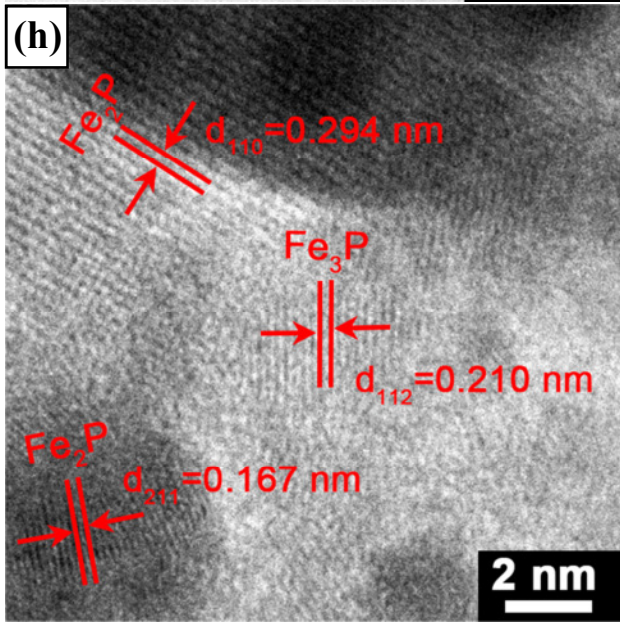
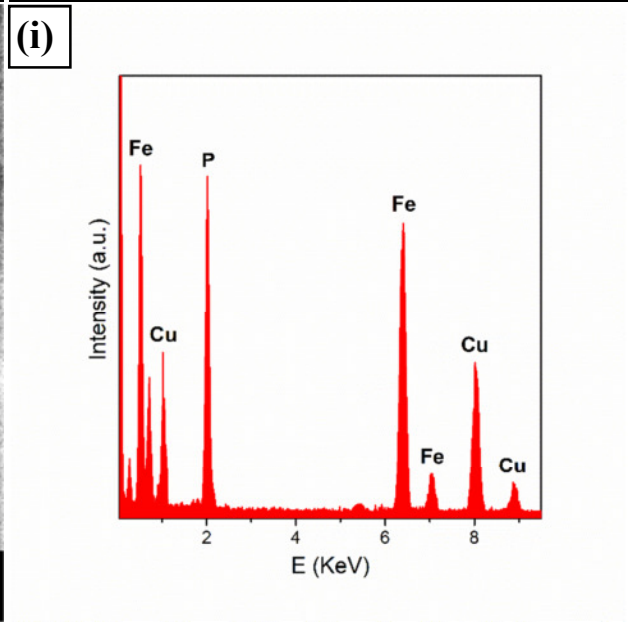

Figure 3

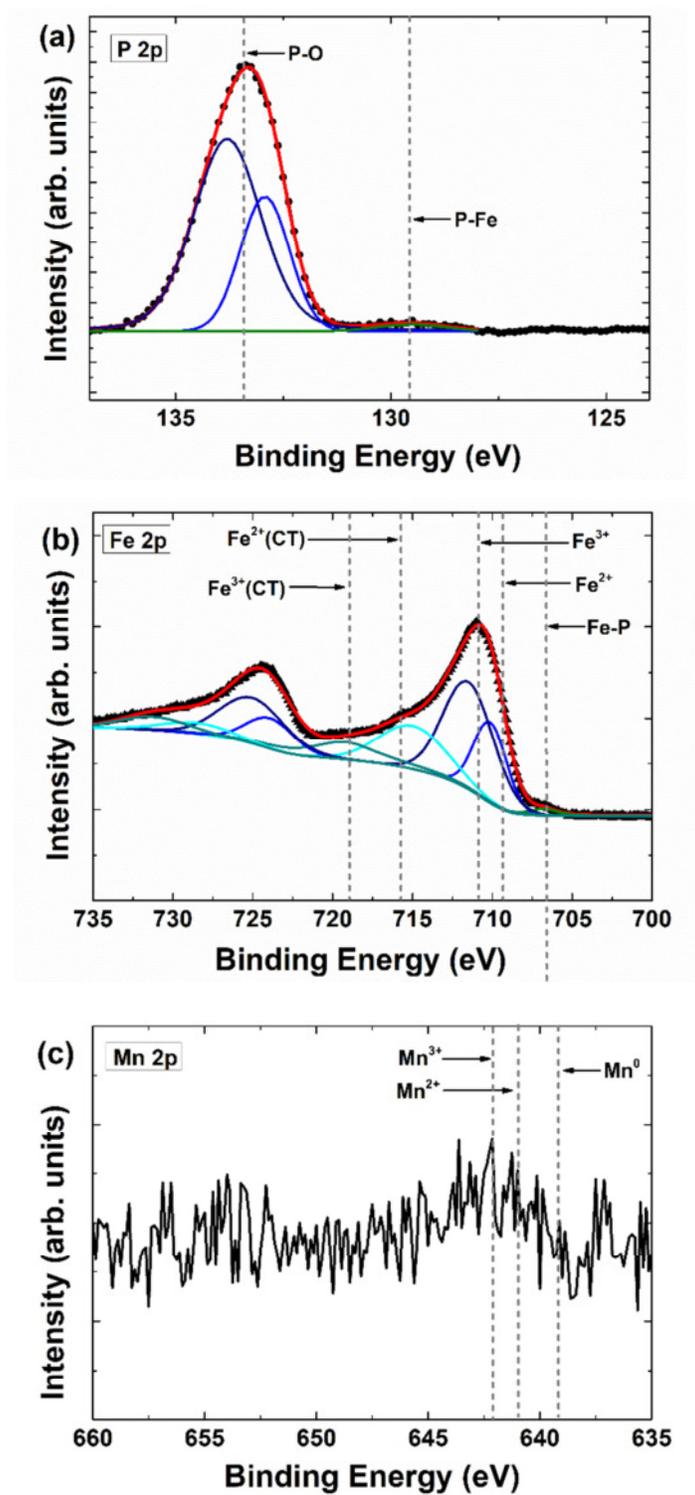

Figure 4

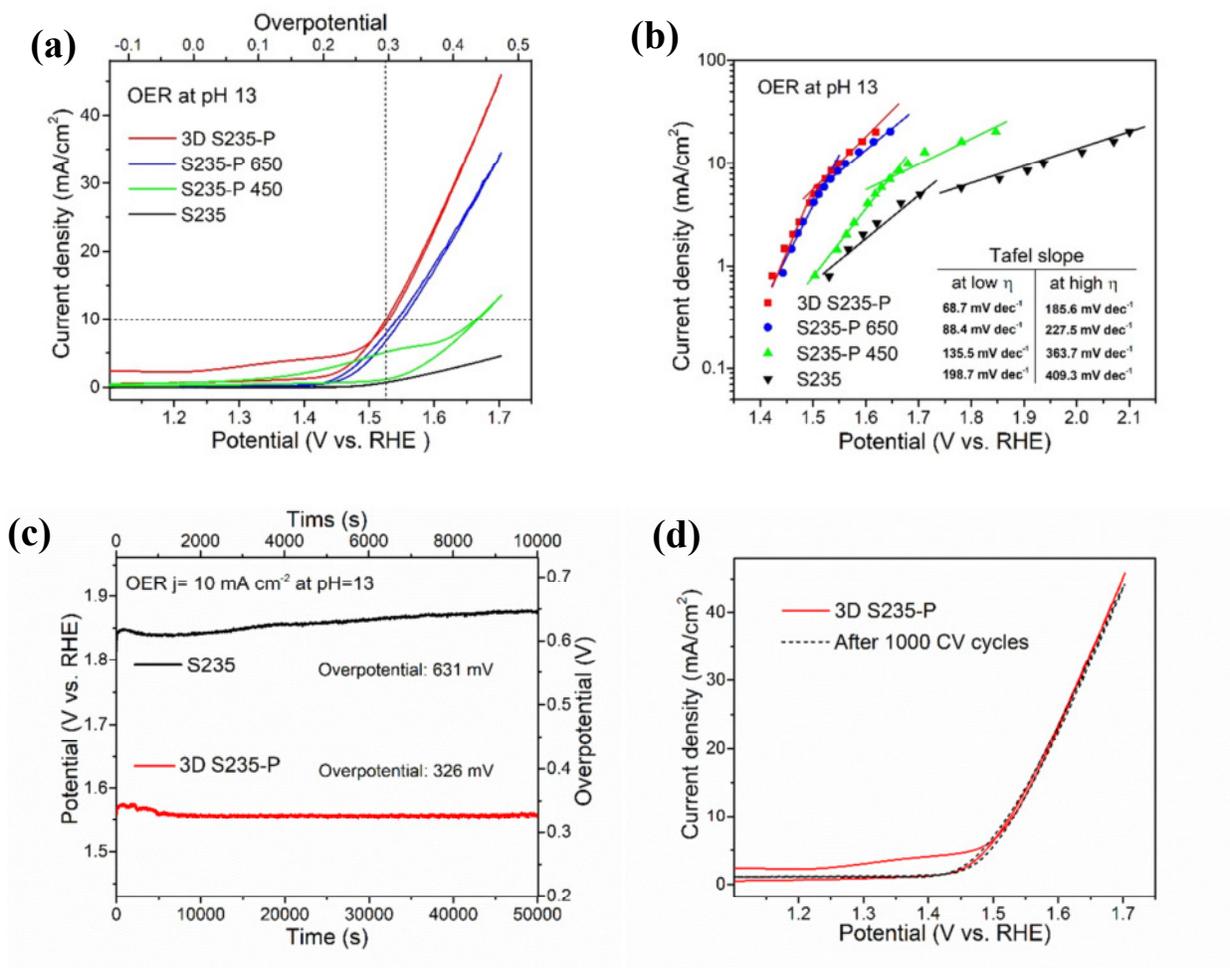

Figure 5

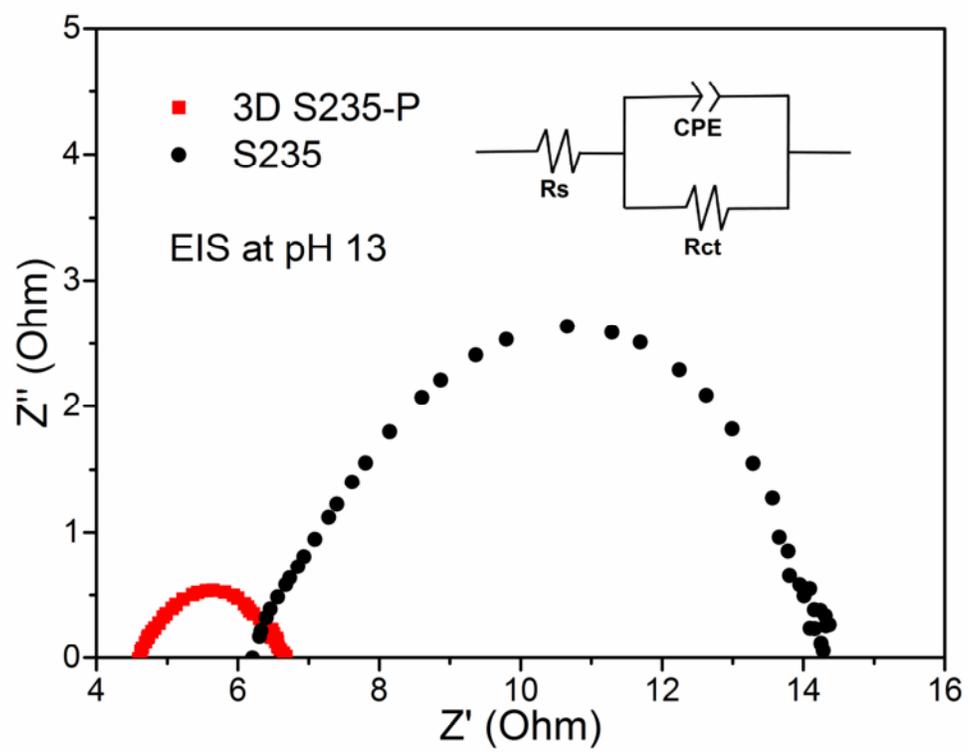

Figure 6

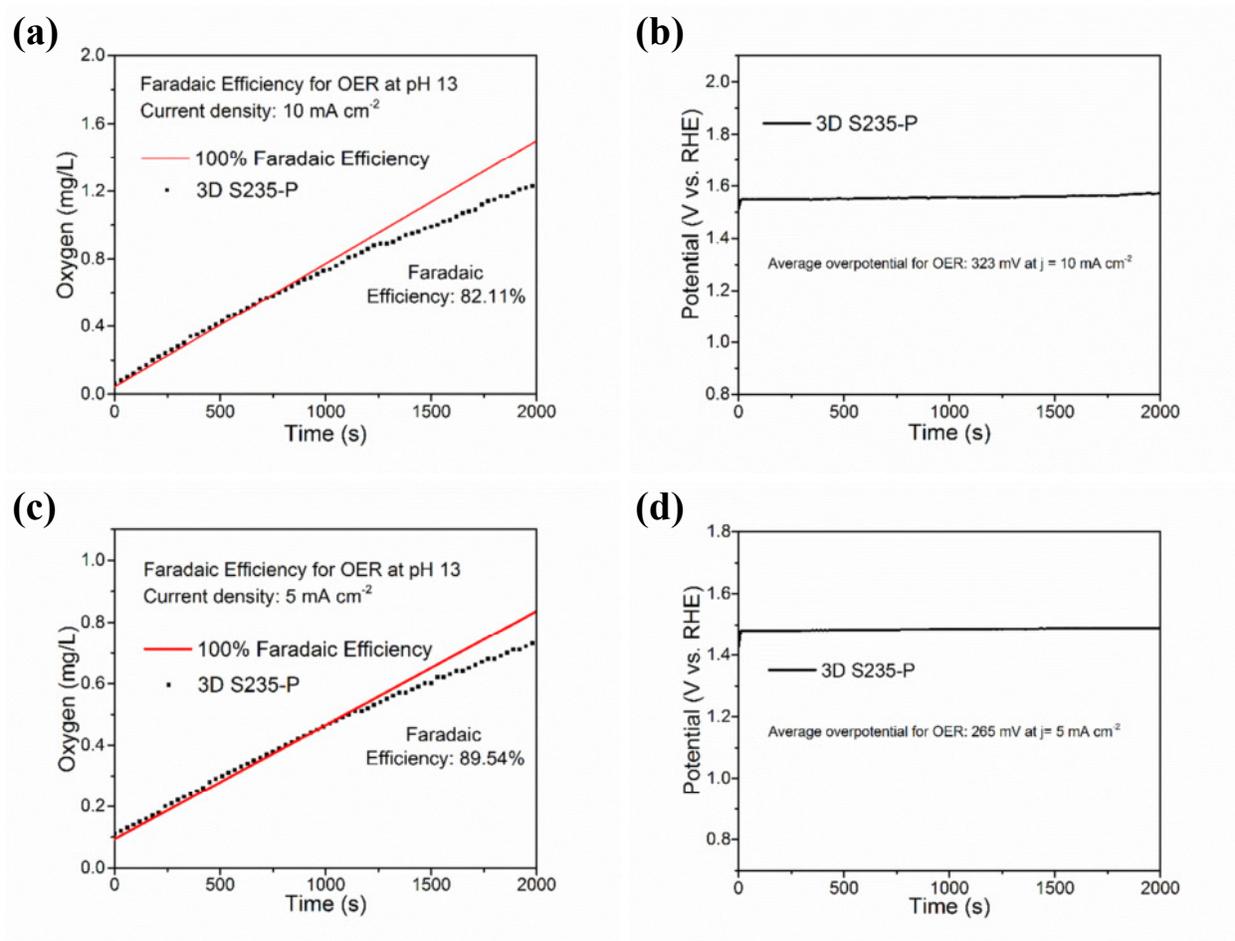

Figure 7

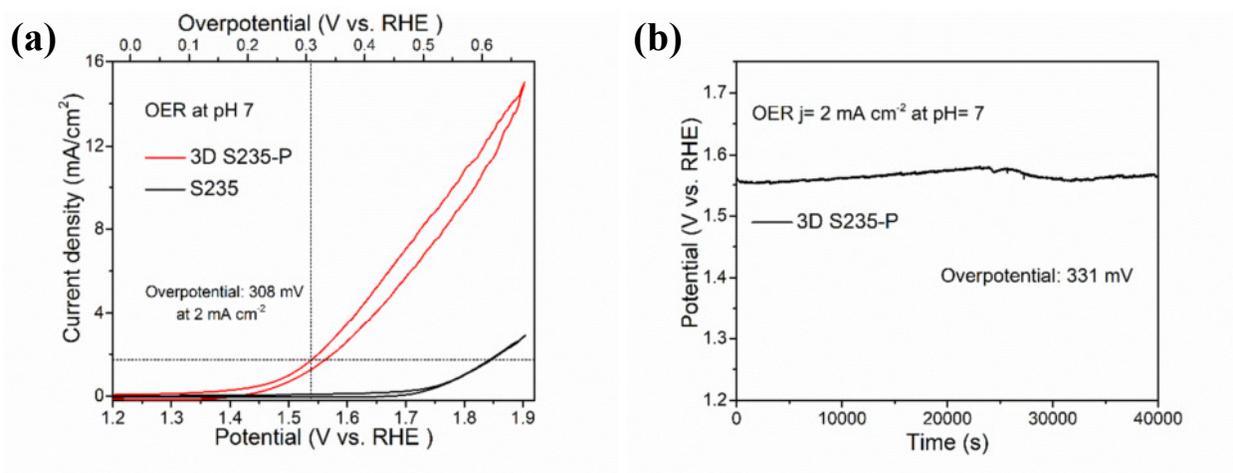

Figure 8

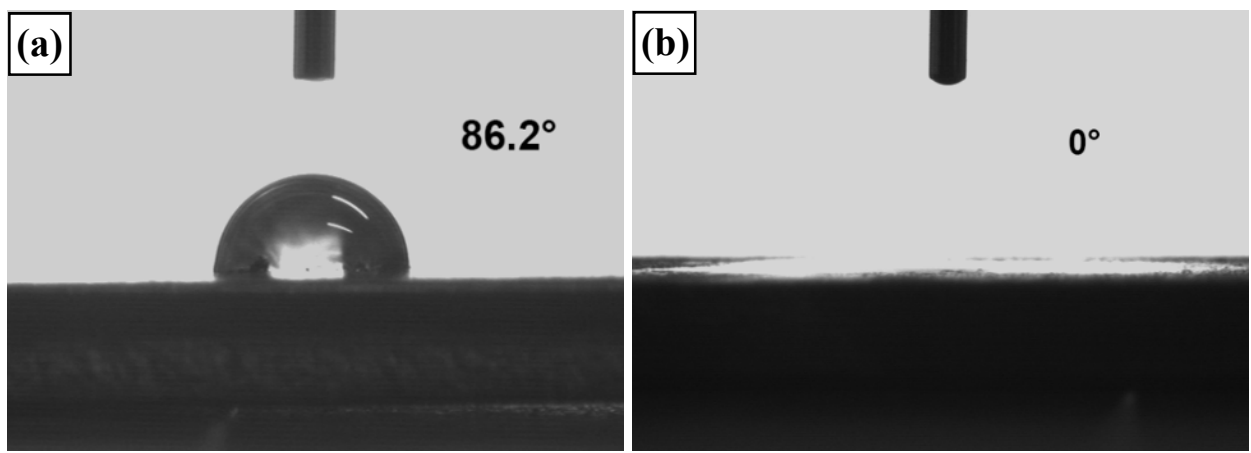
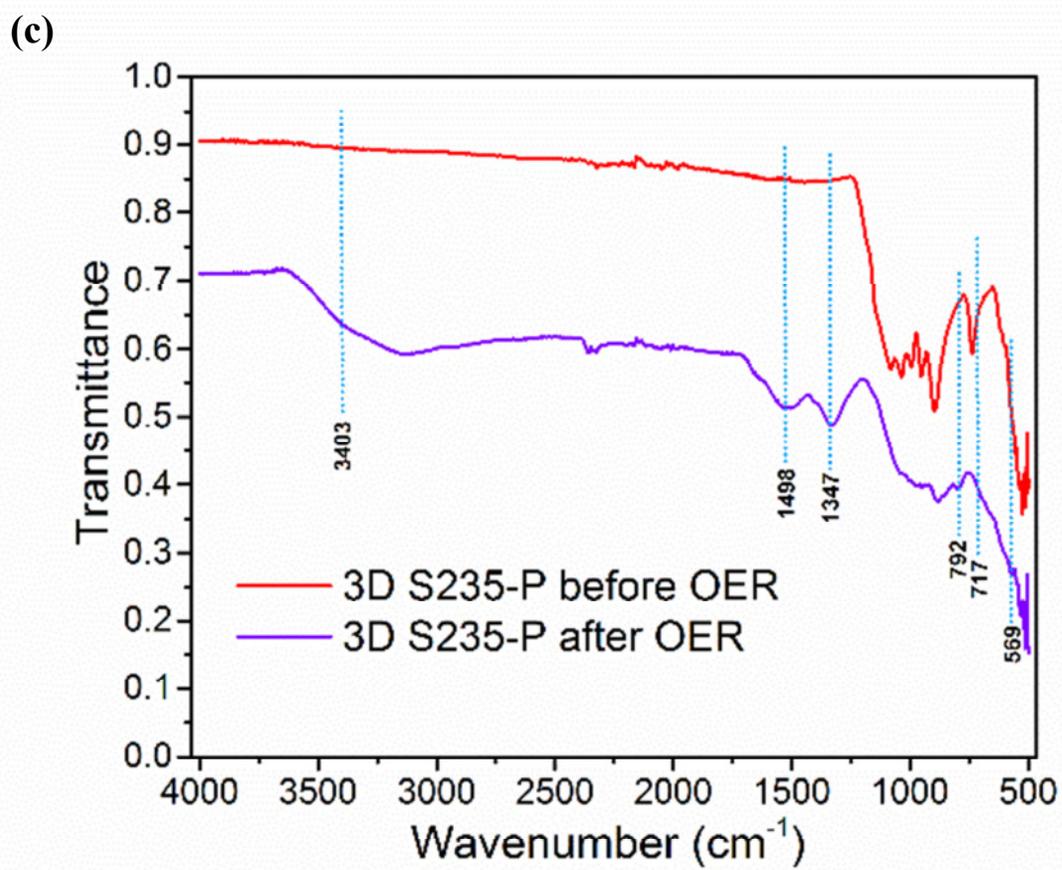

TOC

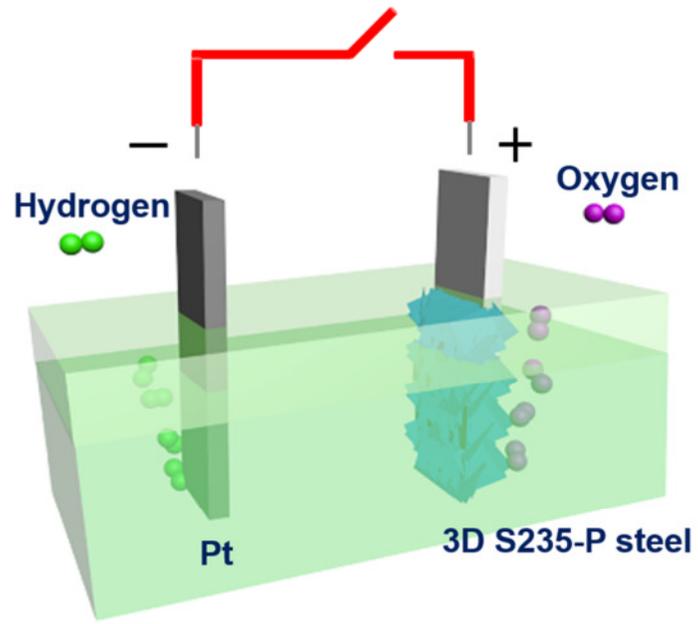